# Electrical transport measurements for superconducting sulfur hydrides using boron-doped diamond electrodes on beveled diamond anvil

*Ryo Matsumoto[a], Mari Einaga[d], Shintaro Adachi[b,e], Sayaka Yamamoto[b,c], Tetsuo Irifune[f], Kensei Terashima[b], Hiroyuki Takeya[b], Yuki Nakamoto[d], Katsuya Shimizu[d], and Yoshihiko Takano[b,c]*

[a]*International Center for Young Scientists (ICYS),*
*National Institute for Materials Science, Tsukuba, Ibaraki 305-0047, Japan*
[b]*International Center for Materials Nanoarchitectonics (MANA),*
*National Institute for Materials Science, Tsukuba, Ibaraki 305-0047, Japan*
[c]*University of Tsukuba, Tsukuba, Ibaraki 305-8577, Japan*
[d]*KYOKUGEN, Graduate School of Engineering Science, Osaka University,*
*Toyonaka, Osaka 560-8531, Japan.*
[e]*Faculty of Engineering, Kyoto University of Advanced Science, Ukyo-ku, Kyoto 615-8577, Japan*
[f]*Geodynamics Research Center, Ehime University, Matsuyama, Ehime 790-8577, Japan*

**Abstract**

A diamond anvil cell (DAC) which can generate extremely high pressure of multi-megabar is promising tool to develop a further physics such a high-transition temperature superconductivity. However, electrical transport measurements, which is one of the most important properties of such functional materials, using the DAC is quite difficult because the sample space is very small and a deformation of electrodes under extreme condition. In this study, we fabricated a boron-doped diamond micro-electrode and an undoped diamond insulation on a beveled culet surface of the diamond anvil. By using the developed DAC, we demonstrated the electrical transport measurements for sulfur hydride $H_2S$ which known as a pressure-induced high-transition temperature superconducting $H_3S$ at high pressure. The measurements were successfully conducted under high pressure up to 192 GPa, and then a multi-step superconducting transition composed from pure sulfur and some kinds of surfer hydrides, which is possible $HS_2$, was observed with zero resistance.

# 1. Introduction

High-pressure techniques strongly support a design of functional materials, for example, hydrogen-storage alloys [1], thermoelectric materials [2-4], superconductors [5-10], and so on [11]. Especially pressure-induced room temperature superconducting materials are expected in hydrogen-rich materials [12-13]. Indeed, the highly compressed sulfur hydride $H_3S$ at 150 GPa exhibits high superconducting transition temperature ($T_c$) above 200 K [14,15]. Much higher $T_c$ was reported in lanthanum hydride under above 170 GPa with record $T_c$ of above 250 K which is very close to room temperature [16,17]. Theoretical prediction suggests that a compressed $Li_2MgH_{16}$ exhibits possible room temperature superconductivity [13].

However, the high-pressure synthesis for the superconducting hydrides needs complex compression-pathway on the temperature-pressure diagram [18], providing a technical difficulty. Moreover, an in-situ electrical transport measurements for an observation zero resistance of their superconductivity is quite difficult. A diamond anvil cell (DAC), which is composed by paired diamond anvil with a culet fabrication, is a unique tool to generate a static high pressure of multi-megabar range [19,20]. If electrodes and insulating layer are inserted into the sample space, the in-situ electrical transport measurements can be performed under high pressure [21]. However, such measurements are quite difficult because the typical diameter of sample space is less than 100 μm, and four electrodes should be inserted into the narrow sample space to measure a sample resistivity. Even if the cell itself is successfully prepared, the electrodes and insulation are sometimes broken by applying pressure.

To solve these problems, a novel configuration of DAC for the electrical transport measurements under high pressure has been developed [22-24]. The combination of a culet-type diamond as a top anvil and box-type nano-polycrystalline diamond (NPD) [25] as a bottom anvil was used in the cell-assembly. The bottom anvil equips micro-electrodes of boron-doped diamond (BDD) [26,27] for the transport measurements, which were fabricated by combining a microwave-assisted plasma chemical vapor deposition (MPCVD) method and an electron beam lithography (EBL) technique [22,23]. Moreover, an undoped diamond (UDD) is also fabricated as an insulating layer on the diamond anvil [24]. The electrical transport measurement can be more easily performed by just putting a sample and gasket on the bottom anvil directly. Since the electrodes and insulating layer are grown from the diamond anvil epitaxially, these components can be repeatedly used until the diamond anvil itself broken. Especially for the synthesis of hydrides, the stable electrodes are significant advantage.

In this study, we successfully fabricated the BDD micro-electrodes and UDD insulating layer on a beveled culet diamond anvil instead of the conventional box-type anvil to achieve higher pressure region and measure the electrical transport properties in the superconducting hydrides more easily. By using the developed DAC, we demonstrated a low-temperature and high-pressure synthesis and in-situ electrical transport measurements for sulfur hydrides $H_2S$ which known as a mother material for high-$T_c$ $H_3S$ [14,15]. A temperature dependence of a sample resistance was measured via a standard four probe method using physical property measurement system (PPMS/Quantum Design). A crystal structure was evaluated using a synchrotron X-ray diffraction (XRD) measurement under high pressure.

## 2. Fabrication for BDD electrodes and UDD insulating layer on beveled anvil

The BDD electrodes and UDD insulating layer were fabricated via following procedures as shown Fig. 1. (a) The shape of the electrodes was drawn by a resist via a lithographic process by EBL in combination with a scanning electron microscopy device (JEOL, JSM-5310) equipped with a nanofabrication system (Tokyo Technology, Beam Draw) on the beveled anvil. (b) The metal mask of Ti/Au thin film was deposited on the resist through a lift-off process. A TiC intermediate layer between the diamond and Ti was formed by annealing at 450°C for 1h in a furnace with Ar gas-flow to enhance adhesion of the metal mask. (c) The epitaxial BDD was selectively grown by a hydrogen-induced MPCVD using $CH_4$ as a source gas onto the uncovered region of the anvil surface. The boron/carbon ratio was tuned to 2500ppm using $C_3H_9B$ gas. The total pressure, total gas flow rate, and microwave power during the growth were maintained at 70 Torr, 300 sccm, and 500 W, respectively. The BDD electrodes were obtained after the wet etching treatment to remove the metal mask and an impurity of graphite using the mixture of $HNO_3$ and $H_2SO_4$ at 400ºC for 30 min. As shown in fig 1 (d), the UDD insulating layer was deposited on around the BDD electrodes using similar processes except for the growth condition of diamond. The total pressure, total gas flow rate of $H_2$ and $CH_4$, and microwave power during the growth were maintained at 35 Torr, 400 sccm, and 350 W, respectively.

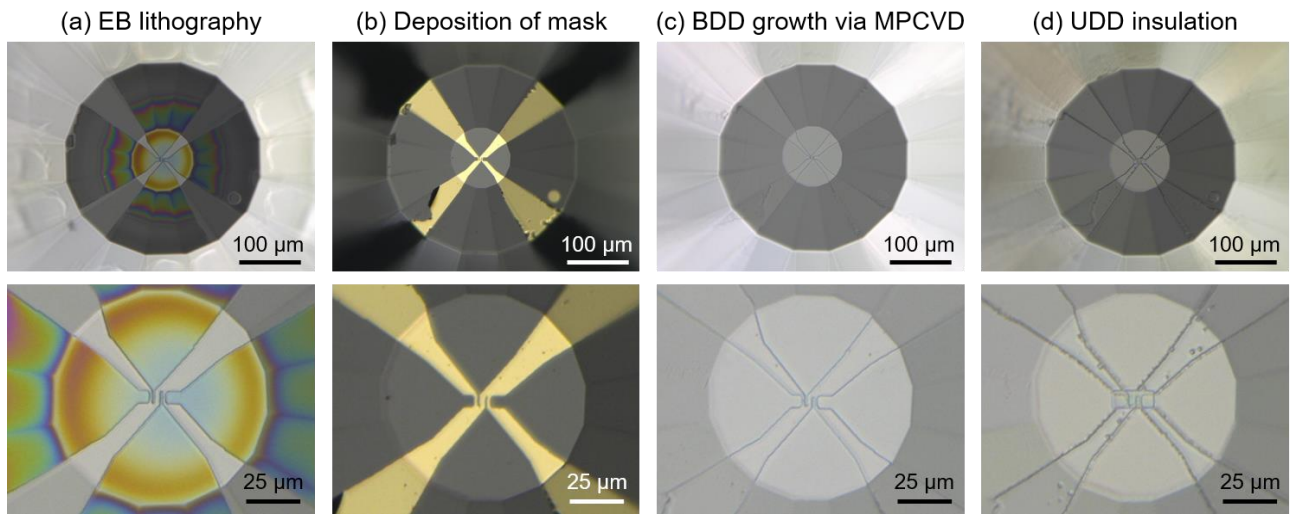

**Figure 1. Photographic images for fabrication processes of BDD electrodes and UDD insulating layer on the culet diamond. (a) lithographic process. (b) fabrication of Ti/Au metal mask. (c) deposition of BDD diamond electrodes. (d) fabrication of UDD insulating layer.**

## 3. Sample preparation

The in-situ electrical transport measurements were performed for sulfur hydrides using the developed DAC. Figure 2 shows schematic and photographic image of DAC with BDD electrodes and UDD insulating layer for a synthesis of superconducting sulfur hydride. The diameter of first anvil and second anvil are 300 μm and 100 μm, respectively. The sample space surrounded by UDD insulation is 15 μm × 20 μm. A gasket was composed by center part of cubic boron nitride (cBN) and other part of rhenium plate. The hole of 30 μm diameter as a sample chamber was fabricated by focused ion beam (Hitachi High-Technologies, SMI9800SE) on the center of cBN part. The pressure value was estimated by the Raman spectrum from the culet of top diamond anvil [28] by an inVia Raman Microscope (RENISHAW).

The sample preparation for sulfur hydrides was performed via a low-temperature and high-pressure pathway by referring the previous report [15,29] of the synthesis for high-$T_c$ $H_3S$. The DAC just before a compression was cooled down by dipping liquid nitrogen and the temperature of the sample space was monitored by a thermocouple. The gas $H_2S$ provided on around the gasket hole, then the gas was cooled down and became solid state. A screw for increasing a stroke length of DAC was rapidly squeezed when the temperature stayed around 200 K and the solid $H_2S$ changed to liquid state, and then the pressure value of the sample space was rapidly increased up to 120 GPa.

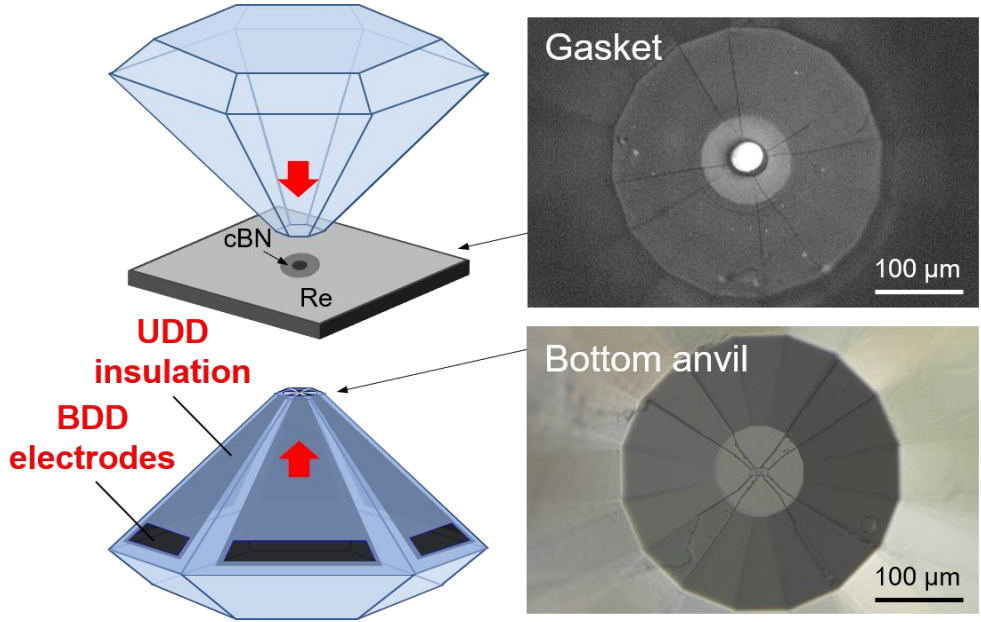

**Figure 2. Schematic and photographic image of DAC with BDD electrodes and UDD insulating layer for a synthesis of superconducting sulfur hydride.**

**4. Electrical transport measurements for sulfur hydrides**

After the temperature of sample space was increased up to room temperature, the electrical resistance measurement was started. Figure 3 (a) shows a time dependence of the resistance of the compressed sulfur hydride, where $t = 0$ means just after the compression from ambient pressure to 120 GPa. The sample resistance was drastically decreased from 77 kΩ to 330 Ω during the pressure evaluation using Raman spectroscopy for approximately 1 hour, indicating a rapid chemical reaction, known as a room temperature annealing effects on sulfur hydrides [14]. The resistance continued to decrease over one night, and then downed to 80 Ω after around 8 hours, which is smaller more than three orders of magnitude compared with that after just compression. An optical microscope image of the sample space as shown in the inset of Fig. 3 (a) exhibits an inhomogeneous morphology including shiny and dark parts.

After 26 hours from the compression the resistance downed to 38 Ω, then we started an evaluation for a temperature dependence of the resistance as shown in Fig. 3 (b) under the pressures up to 154 GPa to examine the superconductivity of the sample. The compressed sample exhibited a semiconducting behavior from 300 K to 30 K and sudden decrease of the resistance from around 25

K at 120 GPa. Corresponding to an increase of the pressure, the sample resistance was continued to decrease up to 154 GPa. Figure 3 (c) shows a magnified plot of the temperature dependence of the resistance from 30 K to 10 K under various pressures. A common feature that the resistance exhibits first drop ($T_{c1}$) from around 25 K and second drop ($T_{c2}$) below 15 K and then achieved to zero was observed under all pressures. Although the $T_{c1}$ is almost independent on the pressure, the $T_{c2}$ is proportionally decreased with an increase of the pressure. It could be suggested that the $T_{c1}$ and $T_{c2}$ were from different superconductivity.

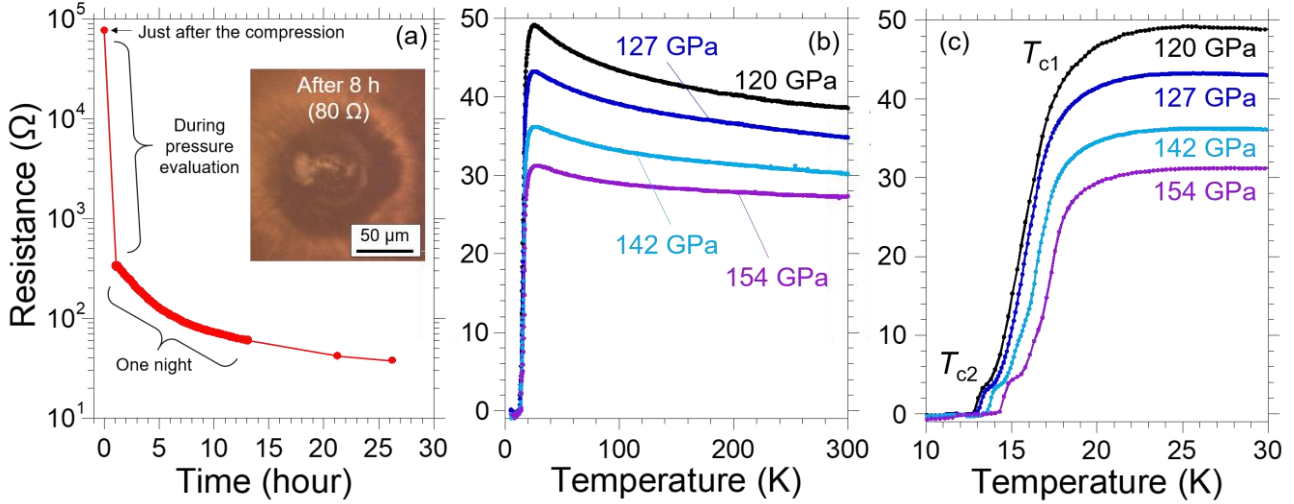

Figure 3. (a) Time dependence of the resistance of the compressed sample, where *t* = 0 means just after the compression from ambient pressure to 120 GPa. (b) Temperature dependence of the resistance under various pressures up to 154 GPa. (c) Magnified plot for the Temperature dependence of the resistance.

To confirm that the two kinds of drops of resistance in the compressed sample are originated from different superconductivity, we measured the temperature dependence of the resistance under magnetic field up to 7 T at 154 GPa, as shown in Fig. 4 (a). The $T_{c1}$ was gradually decreased by an increase of the applied magnetic field. On the other hand, The $T_{c2}$ was rapidly suppressed under the magnetic field and could not be observed above 2 T. To investigate a critical field for $T_{c2}$ phase, we measured the lower magnetic field region as shown in Fig. 4 (b). Figure 4 (c) shows temperature dependence of upper critical field $H_{c2}$ estimated from the Werthamer-Helfand-Hohenberg (WHH) approximation [30] for Type II superconductors in a dirty limit. The extrapolated $H_{c2}(0)$ were 53.0 T and 28.2 T for $T_{c1}$ and $T_{c2}$, respectively, under 154 GPa. From the Ginzburg–Landau (GL) formula $H_{c2}(0) = \Phi_0/2\pi\xi(0)^2$, where the $\Phi_0$ is a fluxoid, the $\xi(0)$ is the coherence length at zero temperature, the obtained values for $\xi(0)$ were 2.5 nm and 3.4 nm for $T_{c1}$ and $T_{c2}$, respectively.

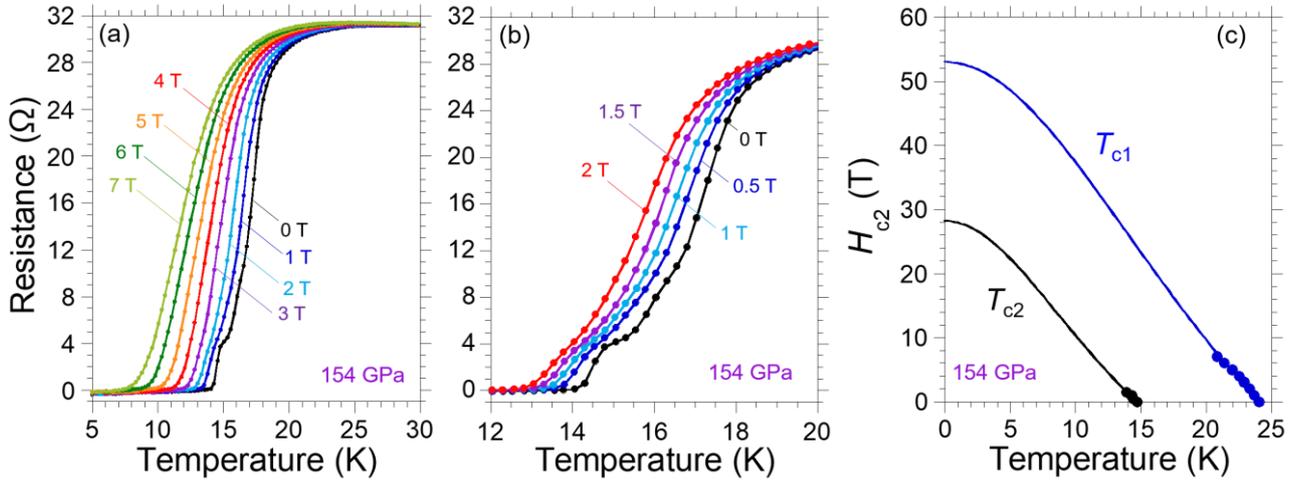

**Figure 4.** (a) Temperature dependence of the resistance at 154 GPa under various magnetic fields up to 7 T. (b) Lower magnetic field region at around $T_{c2}$. (c) Temperature dependence of upper critical field $H_{c2}$ estimated from the WHH approximation.

Here, we note a possible higher onset $T_c$ in the observed superconductivity. Figure 5 shows an enlargement of the temperature dependence of resistance at around $T_{c1}$ under magnetic fields up to 7 T. The separation between the resistances above 30 K and estimated the onset of transition to be at a value of ~35 K. The onset is gradually shifted to lower temperatures with increasing magnetic field. To clear the onset of transition, differential curves of resistance for temperature $dR/dT$ under 0 T and 7 T were shown in the inset of Fig. 5. We can see a clear separation of differential curves from around 35 K under 0 T and 7 T. This is suggestive for an existence of higher $T_c$ region above 35 K in the inhomogeneous sample chamber. The situation is similar to a high-quality boron-doped superconducting diamond [31].

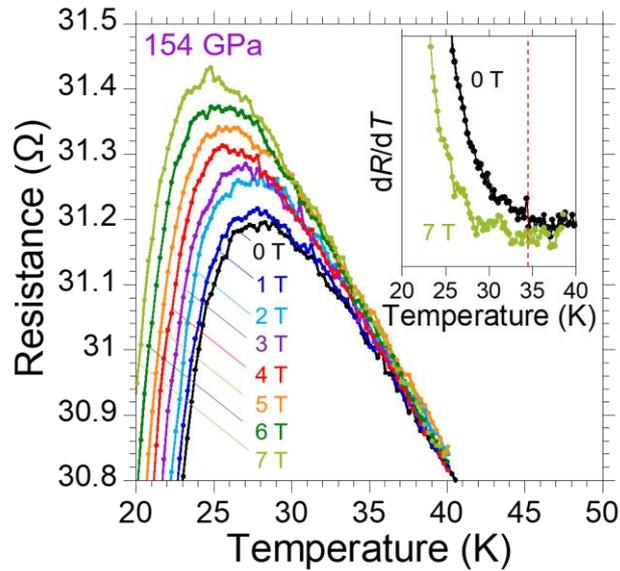

**Figure 5.** Enlargement of temperature dependence of resistance at around $T_{c1}$ under magnetic fields up to 7 T. The inset is differential curve of resistance for temperature.

The sample resistance was continued to decrease slowly even after the compression under 154 GPa. To promote the chemical reactions, we replace the DAC itself onto an oven, and then annealed at various temperature as shown in Fig. 6 (a), where $t = 0$ means just after the *R-T* measurement at 154 GPa. Although the sample exhibited independent behavior against the heating up to 100ºC, the resistance decreased down to around 23 Ω. After the oven heating, we performed a further heating laser and investigated a crystal structure by XRD measurements at SPring-8 (beamline BL10XU). The X-ray beams were monochromatized to a wavelength of 0.41320 Å. The XRD spectra showed no diffraction peak less than 20º, which suggests that the main phase of the produced sample is not bcc $H_3S$ [15,32]. The possible product is the other crystalline sulfur hydrides or some amorphous phases. The IR-laser with wavelength of 1080 nm was radiated by a power of 10-20 W. The total radiation period is around 20 minutes. The obvious differences of the sample resistance and XRD spectrum could not observed after the laser heating. Figure 6 (b) displays the temperature dependence of the resistance from 154 GPa after the heating to 192 GPa in 0-300 K range and fig. 6 (c) 10-30 K range. The sample resistance after the heating was continued to decrease with increasing the pressure value. The $T_{c2}$ increased with the increase of the pressure in contrast to the independent feature of the $T_{c1}$ against the pressure. During the increase of the pressure from 192 GPa, both diamond anvils were broken, unfortunately.

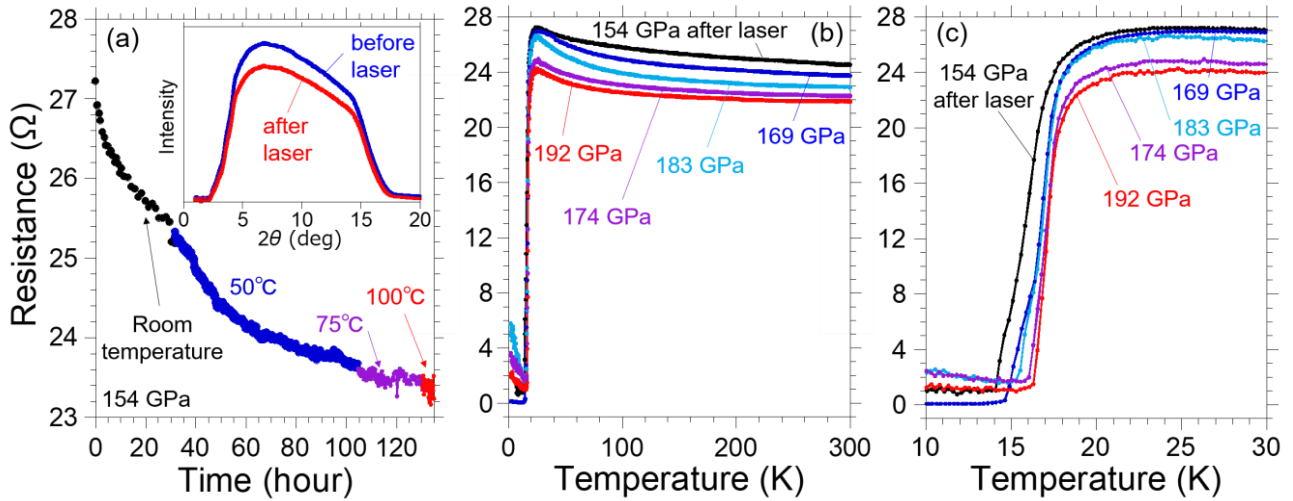

**Figure 6. (a) Time dependence of the resistance at various temperature up to 100ºC, where $t = 0$ means just after the *R-T* measurement at 154 GPa. The inset is XRD pattern for the sample at 154 GPa before and after the laser heating. (b) Temperature dependence of the resistance under various pressures up to 154 GPa. (c) Magnified plot for the Temperature dependence of the resistance.**

Figure 7 displays a comparison of the reported $T_c$ values in sulfur hydrides including both experimental and theoretical studies. The $T_c$ from the experimental works [14,15,33] are shown in solid symbols and the theoretical values [34-38] are shown in open symbols. The criterion of Tc values for our experimental data is shown in a supplemental fig. S1. The pressure dependence of our data of $T_{c2}$ well corresponds to that of the elemental sulfur. The relatively rapid suppression against the applied magnetic fields and the pressure-phase diagram suggest that the $T_{c2}$ from our experiments is originated from the superconductivity in high-pressure phase of elemental sulfur [14,39]. On the other hand, the $T_{c1}$ is not agreement with previously reported $T_c$ from sulfur hydrides, including higher and lower-$T_c$ phase. Among the reported sulfur hydrides, the theoretically calculated pressure-phase diagram for C2/c $HS_2$ is smoothly connected to our data [36], although the $T_{c1}$ was measured up to only 192 GPa. Here our procedures for a sample preparation was slightly different with the previous study [15,32]. Einaga et al. filled a $H_2S$ in the sample chamber at low temperature and increased the pressure up to few GPa to avoid a decomposition to H and S above 25 GPa at room temperature [18]. The DAC was cooled down and kept to 200 K by cryostat and then the pressure increased up to 150 GPa. In our experiment, the pressure was rapidly increased just after the $H_2S$ filling. One of the possibilities for the appearance of hydrogen-poor phase $HS_2$ is the decomposition of $H_2S$ and elimination of hydrogen from sample chamber. Although the pressure-phase diagram suggests that the possible candidate of the product in our experiment, it is necessary to further studies for understanding the origin of the 30 K class high-$T_c$ superconductivity.

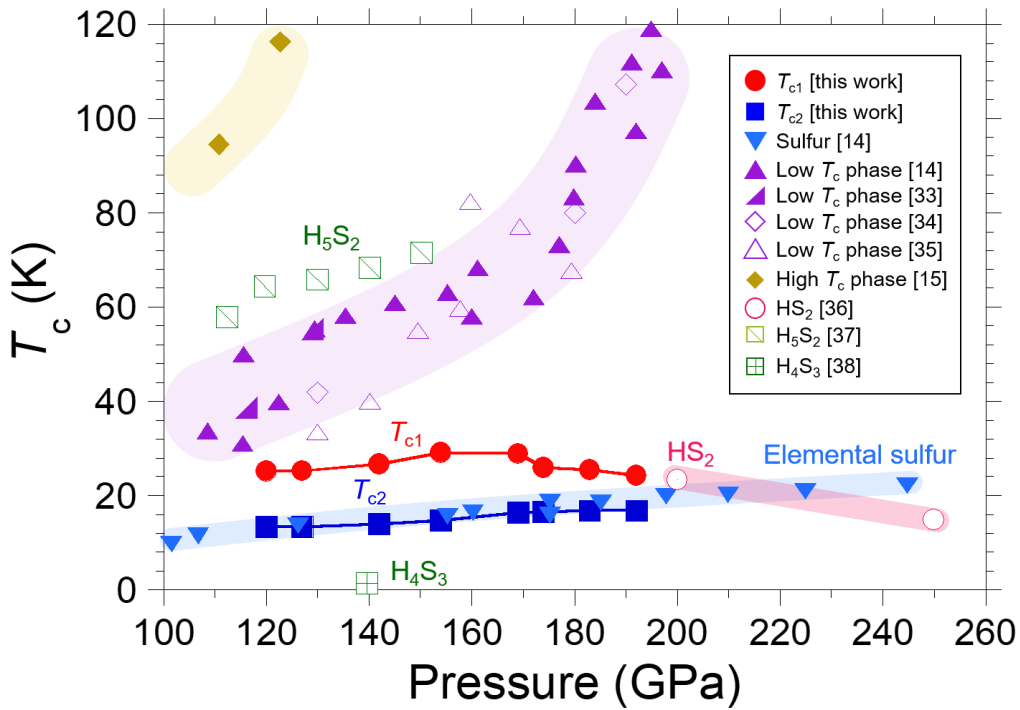

**Figure 7. Comparison of the reported $T_c$ values in sulfur hydrides including both experimental and theoretical studies. The $T_c$ from the experimental works are shown in solid symbols and the theoretical values are shown in open symbols.**

## 4. Conclusion

The BDD micro-electrodes and the UDD insulating layer were successfully fabricated on the beveled culet anvil to perform the in-situ electrical transport measurements under extreme condition above 100 GPa. The mother material $H_2S$ for high-$T_c$ superconducting $H_3S$ was compressed using the developed DAC, and then we observed two-step superconducting transition and clear zero-resistance. From the transport measurements under the magnetic fields and the pressure-phase diagram, we suggest the lower-$T_c$ phase of 15 K class and the higher-$T_c$ phase of 30

K class are originated from elemental sulfur and C2/c HS$_2$, respectively. The developed DAC accelerates to investigate the future high-pressure physics for not only the exploration of superconductors but also wide research fields.

**Acknowledgment**

This work was partly supported by JST CREST Grant No. JPMJCR16Q6, JST-Mirai Program Grant Number JPMJMI17A2, and JSPS KAKENHI Grant Number JP17J05926, 19H02177. A part of the fabrication process of diamond electrodes was supported by NIMS Nanofabrication Platform in Nanotechnology Platform Project sponsored by the Ministry of Education, Culture, Sports, Science and Technology (MEXT), Japan. The part of the high-pressure experiments was supported by the Visiting Researcher's Program of Geodynamics Research Center, Ehime University. The authors would like to acknowledge the ICYS Research Fellowship, NIMS, Japan.

# Supplemental Material for "Electrical transport measurements for superconducting sulfur hydrides using boron-doped diamond electrodes on beveled diamond anvil"

In the plots for our experimental data, the $T_c$s were determined as follows: (1) $T_{c1}$ is average value of $T_{c1}^{lower}$ and $T_{c1}^{upper}$ because the transition is broad. $T_{c1}^{lower}$ was determined from an cross point between the strait line from the normal resistance region under 7 T and extended line from a dropping resistance by superconducting transition under 0 T. $T_{c1}^{upper}$ was determined from a branch point on the d$R$/d$T$ curve under 0 T and 7 T. (2) $T_{c2}$ was determined from an cross point between the strait line from the normal resistance region and extended line from a dropping resistance by superconducting transition. The example for the criterion of $T_{c1}$ and $T_{c2}$ under 120 GPa is shown in fig. S1.

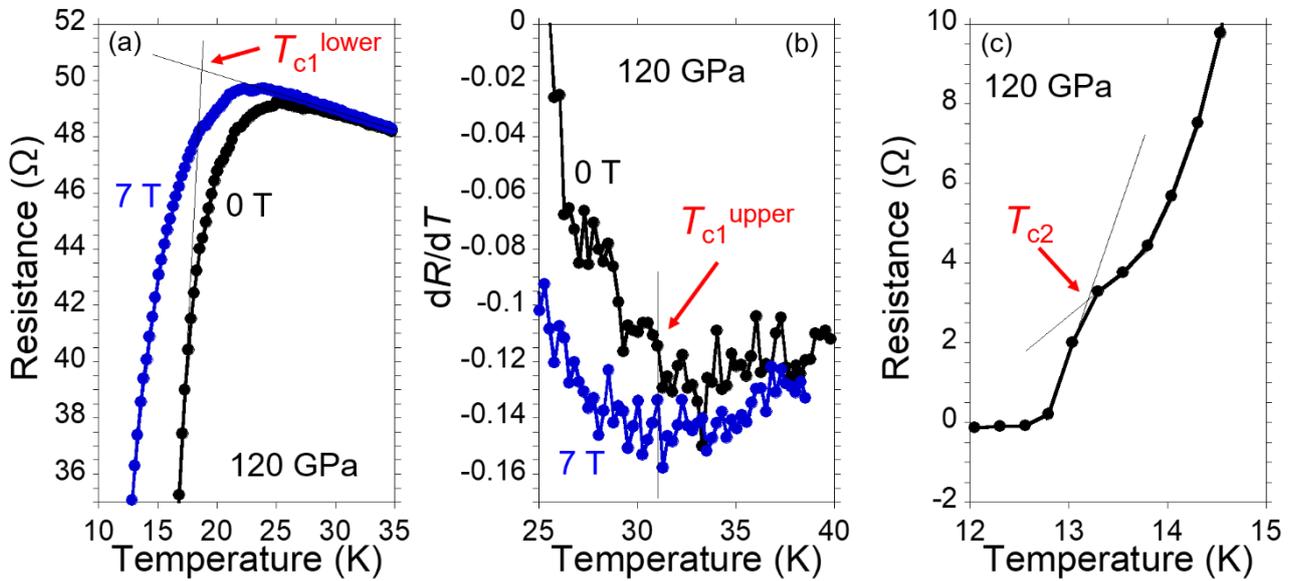

**Figure S1.** The example for the criterion of $T_c$ value for our data in fig. 7. (a) $T_{c1}^{lower}$, (b) $T_{c1}^{upper}$ and (c) $T_{c2}$ under 120 GPa.